\documentstyle[12pt,titlepage]{article}

\advance\textheight by \topskip

\newcommand{\beq}{\begin{equation}}
\newcommand{\eeq}{\end{equation}}
\newcommand{\bea}{\begin{eqnarray}}
\newcommand{\eea}{\end{eqnarray}}
\newcommand{\beas}{\begin{eqnarray*}}
\newcommand{\eeas}{\end{eqnarray*}}
\newcommand{\beqs}{\begin{displaymath}}  
\newcommand{\eeqs}{\end{displaymath}}    

\def\G{{\rm G}}        

\begin{document}

\author{
  O.~v.~Styp Rekowski
  and H.~Frommert
  \\
  \rm E-Mail: \tt Hartmut.Frommert@uni-konstanz.de\\[1ex]
  \em Dept.\ of Physics, University of Constance\\
  \em P.O.Box 55 60 M 678, D-78434 Konstanz, Germany
}

\title{\bf Solution of the Higgs scalar-tensor theory without Higgs particles
  for static stars}

\maketitle

\thispagestyle{empty}
\section*{Abstract}

Within the scalar-tensor theory of gravity with
Higgs mechanism without Higgs particles, we prove that the excited
Higgs potential (the scalar field) vanishs inside and outside of the stellar
matter for static spherically symmetric configurations. The field equation
for the metric (the tensorial gravitational field) turns out to be
essentially the Einsteinian one.

\vfill

\noindent
\begin{description}
\item[Keywords]:
  Higgs scalar-tensor theory;
  Higgs mechanism without Higgs particles;
  particle theoretical implications to gravitation theory;
  static and spherically symmetric solutions
\end{description}

\clearpage
\pagenumbering{arabic}

\section{Introduction}

A scalar-tensor theory of gravity was developed by Brans and Dicke 1961
in order to introduce some foundation for the inertial mass as well as
the active and passive gravitational mass (i.e., the gravitational `constant'),
by a scalar function determined by the distribution of all other particles
in the universe; the background of this is Mach's principle and the principle
of equivalence.

This introduction of mass by a scalar field can now be regarded as a somehow
prophetic approach, because in today's Standard Model of particle physics
the masses of the elementary particles are generated via the Higgs mechanism,
thus using also a scalar field, the Higgs field.
The scalar interaction mediated by the Higgs field was investigated by
Dehnen, Frommert, and Ghaboussi 1990.
They showed that any excited Higgs field%
  \footnote{
    The quanta of this excited Higgs field are the hypothetical Higgs
    particles.
  }
mediates an attractive scalar interaction%
  \footnote{
    This interaction is similar to gravity because it couples to the masses
    of the particles.
  }
of Yukawa type (i.e.\ short range)
between those particles which acquire mass by the corresponding symmetry
breaking (i.e.\ the fermions and the massive $W$ and $Z$ gauge bosons).
The Higgs field of particle physics can also serve as the scalar field in
a scalar-tensor theory of gravity, as was first proposed by Zee 1979 and
deeper investigated by Dehnen, Frommert, and Ghaboussi 1992.
In this theory, in addition to its role in the Standard Model to make the
particles massive, the scalar Higgs field also generates the gravitational
constant G.
Surprisingly however, if the Higgs field of the
$SU(3)\times SU(2)\times U(1)$ Standard Model of the elementary particles
is employed to generate G, the Higgs field looses its source,
i.e.\ can no longer be generated by fermions and gauge bosons
unless in the very weak gravitational channel.

The reader can find the whole formalism of this theory in Dehnen and
Frommert 1993.

\section{Static spherically symmetric solutions of the Higgs scalar-tensor
  theory}

For the excited Higgs field $\varphi$, one obtains
the following homogeneous, covariant Klein-Gordon equation%
  \footnote{
    Throughout this paper we use
    $\hbar=c=1$ and the metric signature $(+---)$.
    The symbol $(\ldots)_{|\mu}$ denotes the partial,
    $(\ldots)_{\|\mu}$ the covariant derivative with respect to the
    coordinate $x^\mu$.
  }
(see Dehnen, Frommert 1993):
\beq
\xi^{|\mu}_{\ \ \|\mu} + M^2 \xi = 0\ ,\ \ \ \xi = (1+\varphi)^2-1\ ,
\label{eq:scf}
\eeq
where $M$ denotes the mass of the Higgs particles in this theory.  The field
equation for the metric as the tensorial gravitational field reads:
\bea
R_{\mu\nu} - {1 \over 2} R g_{\mu\nu} 
&=& - {8\pi\G\over1+\xi} \Biggl[T_{\mu\nu}
+ {v^2\over4\left(1+\xi\right)} \left(\xi_{\vert\mu}\xi_{\vert\nu}
- {1\over2}\xi_{\vert\lambda}\xi^{\vert\lambda} g_{\mu\nu}\right)
+ V(\xi) g_{\mu\nu}\Biggr]
\nonumber \\*
&&{} - {1 \over 1+\xi} \left[\xi_{\vert\mu\Vert\nu}
- \xi^{\vert\lambda}_{\ \ \Vert\lambda} g_{\mu\nu} \right]\ .
\label{eq:gravf3chi}
\eea
with the Ricci tensor $R_{\mu\nu}$
and the Higgs potential
\beq
V(\xi) = {3\over32\pi\G} M^2 \left(1+{4\pi\over3\alpha}\right) \xi^2
  \approx {3 M^2\over32\pi\G} \xi^2\ \ \ \ (\alpha\simeq 10^{33})\ .
\label{eq:pot}
\eeq
$T_{\mu\nu}$ is the energy-momentum tensor of matter.

We now look for the exact solution of this equation for the spherically
symmetric and time independent case. This means that the excited Higgs field
is a function of the radius $r$ only, and the metric has the form
\beq
g_{\mu\nu} =
  \left[
  \begin{array}{cccc}
  e^{\nu(r)} & 0                & 0     & 0 \\
  0          & - e^{\lambda(r)} & 0     & 0 \\
  0          & 0                & - r^2 & 0 \\
  0          & 0                & 0     & - r^2 \sin^2\vartheta
  \end{array}
  \right]\ .
\label{eq:met}
\eeq

Using the Christoffel symbols and the Ricci tensor components
following from the metric (\ref{eq:met})
(see e.g.\ Landau and Lifshitz 1992, \S100, or Tolman 1934, \S98),
the nontrivial field equations for the metric read
(primes denote derivatives with respect to the radial coordinate $r$,
$L=1/M$ the Compton wavelength corresponding to the Higgs mass $M$):
\bea
R_{00} &=&
  - e^{\nu-\lambda} \left(
  {\nu^{\prime\prime}\over 2} + {{\nu^\prime}^2\over 4}
  - {\nu^\prime\lambda^\prime\over 4} + {\nu^\prime\over r}
                  \right)
\nonumber
\\
&=& - {e^{\nu-\lambda}\over 1+\xi} \left[
  4\pi\G\left(\rho+3p\right) e^\lambda
  - {\nu^\prime\xi^\prime\over 2}
  + {\xi\over L^2} \left(1-{3\over 4}\xi\right) e^\lambda
                     \right]
\\
R_{11} &=&
  {\nu^{\prime\prime}\over 2} + {{\nu^\prime}^2\over 4}
  - {\nu^\prime\lambda^\prime\over 4} - {\lambda^\prime\over r}
\nonumber
\\
&=& {1\over 1+\xi} \left[
  - 4\pi\G\left(\rho-p\right) e^\lambda - \xi^{\prime\prime}
  + {\lambda^\prime\xi^\prime\over 2}
  + {\xi\over L^2} \left(1-{3\over 4}\xi\right) e^\lambda
                   \right]
\\
R_{22} &=& e^{-\lambda} - 1
  + {r\over2} \left(\nu^\prime-\lambda^\prime\right) e^{-\lambda}
\nonumber
\\
&=& - {1\over 1+\xi} \left[
  4\pi\G\left(\rho-p\right) r^2 + r \xi^\prime e^{-\lambda}
  - {\xi r^2\over L^2} \left(1-{3\over 4}\xi\right)
                   \right]
\label{eq:R22}
\eea
and the scalar field equation (\ref{eq:scf}) takes the form
\beq
{d^ 2\xi(r)\over dr^ 2} + 
  \left\{{2\over r}+{1\over2}{d\over dr}\left[\nu(r)-\lambda(r)\right]\right\}
     {d\xi(r)\over dr}
  = M^2 e^{\lambda(r)} \xi(r)\ .
\label{eq:fex}
\eeq

\noindent
Because of a continuous and finite matter density, i.e.\ no singularities
such as matter points or infinitely thin massive surfaces, we are looking for
an exact solution for $\xi(r)$ of this equation, which is finite and
continuous together with its first derivative.

We can immediately find the exact solution of equation (\ref{eq:fex}) if the 
metric is the Minkowskian one (perhaps with some constant coordinate 
transformation). This should be a good approximation for the limit of
large distances from the star ($r\gg R$, where $R$ is the radius of the star)
in the static case.
Equation (\ref{eq:fex}) then gets
linearized and becomes the usual Klein-Gordon equation
for a static, spherically symmetric field:
\beq
{d^ 2\xi(r)\over dr^ 2} + {2\over r} {d\xi(r)\over dr} - M^2 \xi(r) = 0
\label{eq:kg}
\eeq
The bounded solution of this equation is the Yukawa function
\beq
\xi(r)={A e^ {-r/L} \over r}\ ,\ \ \ r\gg R\ ,
\label{eq:yuk}
\eeq
with $A$ an arbitrary real constant; this is the asymptotic solution for
all finite spherically symmetric systems for large values of $r$ which are
asymptotically embedded in flat Minkowski spacetime. The absolute value of
this solution is exponentially {\em decreasing\/} as
$r\longrightarrow\infty$.

On the other hand, the spacetime metric is also asymptotically equivalent to
the flat Minkowskian one for the limiting case%
\footnote{
  This follows immediately from the requirement that for our spherically
  symmetric configuration the fields should be differentiable, if one takes
  a look on an arbitrary straight line through the origin: As our fields
  $\nu$, $\lambda$, and $\xi$
  must be spherically symmetric, they must be even functions of the distance
  from the origin on this line, and thus have vanishing derivatives at $r=0$,
  which makes the connection coefficients vanish.

  It also follows as the limiting case of a corollary based on Birkhoff's
  theorem, that the metric inside an empty central spherical cavity
  (of radius $R_i$) in a spherically symmetric system is equivalent to the
  flat Minkowski metric, for $R_i\longrightarrow 0$. This corrolary is
  treated e.g.\ in Weinberg 1972, and is also valid in our scalar-tensor
  theory.
}
$r\longrightarrow0$. Therefore, the scalar field near $r=0$ should be given
asymptotically again by a solution of equation (\ref{eq:kg}); in this case,
the solution should behave regular at $r=0$ to avoid singularities.
The regular solution at $r=0$ of (\ref{eq:kg}) is given by
\beq
\xi(r)={B\sinh(r/L)\over r}\ ,\ \ \ 0\leq r \ll R
\label{eq:sinh}
\eeq
($B$ another arbitrary real constant),
the absolute value of which has a minimum at $r=0$ and is {\em increasing\/}
outward.

In addition, we can discuss the limiting case for small values of $r$ more
acurately:
For the interior solution near the origin at $r=0$, it is convenient to
rewrite the field equation (\ref{eq:fex}) after multiplication with $r$:
\beq
{r\over2} \xi^{\prime\prime} +
  \left[1+{r\over4}\left(\nu^\prime-\lambda^\prime\right)\right] \xi^\prime
  = {r\over2} M^2 e^{\lambda(r)} \xi\ .
\label{eq:rfex}
\eeq
Obviously, for nonsingular fields, $\xi^\prime(r)$ must vanish at $r=0$.
Taylor-expanding $\xi(r)$ as $\xi(r)=\xi_0+\xi_1 r+\xi_2 r^2+\ldots$ yields
\bea
\xi_1 &=& 0
\\
\xi_2 &=& {M^2\over 6} e^{\lambda_0} \xi_0\ ,\ \ (\lambda_0 = \lambda(r=0))
\eea
which shows that the second derivative $\xi^{\prime\prime}$ of the scalar
field $\xi$ has the same sign at the origin $r=0$ as $\xi(r=0)$.

If $\xi_0=\xi(r=0)$ is not zero, i.e.\ $\xi$ doesn't vanish identically,
its absolute value anyway increases outward from the center, i.e.\ if
$\xi_0$ is positive, $\xi$ increases, and if it is negative, it deceases
outward.
One could expect that the complete {\em exact solution} of equation
(\ref{eq:fex})
would have a maximum for every $A>0$ in equation (\ref{eq:yuk})
or $B>0$ in (\ref{eq:sinh}), because it grows when
starting from $r=0$, and vanishes exponentially as $r\longrightarrow\infty$.
On the other hand, its
first derivative would vanish at this extremal point, and then equation 
(\ref{eq:fex}) would force the same sign on the solution $\xi(r)$ and its
second derivative.
Because the function $\xi(r)$ is positive, one would obtain a minimum
and not a maximum at this point.
For $A<0$ or $B<0$, we have the analogous situation: one expects
at least one minimum and gets only a maximum.
Therefore, one cannot get the assymptotically bounded exterior solution
(\ref{eq:yuk}) from any nontrivial solution which behaves regular near 
r=0.
Thus the only physically permitted static solution is $\xi(r)\equiv0$,
with the constants $A=0$ and $B=0$ for the asymptotic solutions.

\section{Conclusions}

We have shown that the only physically permitted solution for a static,
spherically symmetric configuration in our theory is the trivial one
with respect to the scalar field.
Therefore, the gravitational tensor field equation becomes an ordinary
Einstein equation, so that all calculations for the astronomical objects
obtained from Einstein's General Relativity stay valid.
Of course, this approach is only manifest for the exactly
spherically symmetric and static case without pointlike singularities,
and it does not cover highly dynamic systems
(e.g., cosmological models or black holes).
Yet it is a good approximation for a great deal of ``normal''
objects like stars, or perhaps all closed systems, e.g. our Solar system;
for all these our fundamental result should be valid.

As the physical world is dynamic, however, there remains the 
possibility of dynamic solutions which assymptotically fit to a 
cosmological background (see e.g.\ Frommert, Schoor, and Dehnen 1996).
This may be of interest in the context of the dark matter problem.

\section*{Acknowledgements}

The authors are thankful to
Heinz Dehnen, Sokratis Rahoutis, and Holger Schoor
for helpful hints and discussions.


\end{document}